\documentclass[fleqn,usenatbib]{mnras}
\usepackage{pbox,units}
\usepackage[flushleft]{threeparttable}
\usepackage[normalem]{ulem}
\usepackage{newtxtext,newtxmath}

\usepackage[T1]{fontenc}
\usepackage{ae,aecompl}

\usepackage{graphicx}	
\usepackage{amsmath}	
\usepackage{amssymb}	
\usepackage{multirow}

\title[The minimum and maximum GWB for PTAs]{The minimum and maximum gravitational-wave background from supermassive binary black holes}

\author[Zhu, Cui \& Thrane]{
Xing-Jiang Zhu$^{1,2}$\thanks{E-mail: xingjiang.zhu@monash.edu}, Weiguang Cui$^{3}$ and Eric Thrane$^{1,2}$
\\
$^{1}$School of Physics and Astronomy, Monash University, Clayton, Vic 3800, Australia\\
$^{2}$OzGrav: The Australian Research Council Centre of Excellence for Gravitational Wave Discovery\\
$^{3}$Departamento de F\'isica Te\'{o}rica, M\'{o}dulo 15, Facultad de Ciencias, Universidad Aut\'{o}noma de Madrid, 28049 Madrid, Spain\\
}

\date{Accepted XXX. Received YYY; in original form ZZZ}

\pubyear{2018}

\begin{document}
\label{firstpage}
\pagerange{\pageref{firstpage}--\pageref{lastpage}}
\maketitle

\begin{abstract}
The gravitational-wave background from supermassive binary black holes (SMBBHs) has yet to be detected.
This has led to speculations as to whether current pulsar timing array limits are in tension with theoretical predictions.
In this paper, we use electromagnetic observations to constrain the SMBBH background from above and below.
To derive the {\em maximum} amplitude of the background, we suppose that equal-mass SMBBH mergers fully account for the local black hole number density.
This yields a maximum characteristic signal amplitude at a period of one year $A_{\rm{yr}}<2.4\times 10^{-15}$, which is comparable to the pulsar timing limits.
To derive the {\em minimum} amplitude requires an electromagnetic observation of an SMBBH.
While a number of candidates have been put forward, there are no universally-accepted electromagnetic detections in the nanohertz band.
We show the candidate 3C 66B implies a lower limit, which is inconsistent with limits from pulsar timing, casting doubt on its binary nature.
Alternatively, if the parameters of OJ 287 are known accurately, then $A_{\rm{yr}}> 6.1\times 10^{-17}$ at 95\% confidence level.
If one of the current candidates can be established as a bona fide SMBBH, it will immediately imply an astrophysically interesting lower limit.
\end{abstract}

\begin{keywords}
gravitational waves -- galaxies: evolution --  quasars: individual(OJ 287) --  galaxies: individual(3C 66B) -- black hole physics -- pulsars: general
\end{keywords}



\section{Introduction}
The detection of gravitational waves (GWs) from several stellar-mass compact binary merger events \citep{O1BBH16,GW170817,GW170608} by ground-based laser interferometers Advanced LIGO \citep{aLIGO} and Advanced Virgo \citep{aVirgo} heralded a new era of GW astronomy.
In parallel to ground-based detectors that are monitoring the audio band ($\sim 10-10^{3}$ Hz) of the GW spectrum, decades of efforts have gone into opening the nanohertz frequency ($10^{-9}-10^{-6}$ Hz) window \citep{Sazhin1978,Detweiler1979,Hellings_Downs83,Jenet05,IPTA,IPTAdick13}. These experiments, called pulsar timing arrays \citep[PTAs;][]{Foster_Backer90}, use a number of ultra stable millisecond pulsars collectively as a Galactic-scale GW detector.

Previous PTA searches \citep[e.g.,][]{YardleySGWB,EPTAlimit,NANOGrav2012} have primarily targeted a GW background (GWB) formed by a cosmic population of supermassive binary black holes \citep[SMBBHs;][]{Begelman80}. Assuming that binaries are circular with the orbital evolution driven by GW emission, the characteristic amplitude $h_c$ of such a background signal can be well described by a power-law spectrum $h_{c}\sim f^{-2/3}$ with $f$ being GW frequency \citep{Phinney01}; the amplitude at $f=1\,{\rm{yr}}^{-1}$, denoted by $A_{\rm{yr}}$, can be conveniently used for comparing steadily-improving experimental limits and different model predictions. Prior to the formation of timing arrays, \citet{Kaspi94} used long-term observations of two pulsars to constrain $A_{\rm{yr}}\leq 2.6\times 10^{-14}$ with 95\% confidence (the same confidence level applied for limits quoted below). \citet{Jenet2006} reduced this limit to $1.1\times 10^{-14}$ using a prototype data set of the Parkes Pulsar Timing Array \citep[PPTA;][]{PPTA2013}. Since the establishment of three major PTAs, including PPTA, NANOGrav \citep{NANOGrav} and the European PTA \citep[EPTA;][]{EPTA}, the constraints have been improved by an order of magnitude over the last decade. Currently the best published upper limit on $A_{\rm{yr}}$ is $1\times 10^{-15}$ by the PPTA \citep{PPTA15Sci} with comparable results from the other two PTAs \citep{EPTA15GWBlimit,NANOGrav11GWBlimit}. The three PTAs have joined together to form the International Pulsar Timing Array \citep[IPTA;][]{IPTA16}, aiming at a more sensitive data set.

Over the past two decades, theoretical predictions of $A_{\rm{yr}}$ have also evolved. \citet{RajaRomani95} considered several mechanisms that can drive SMBBHs into the GW emission regime and obtained an estimate of $A_{\rm{yr}}= 2.2\times 10^{-16}$. \citet{Jaffe_Backer03} used galaxy merger rate estimates and the scaling relation between black hole mass and the spheroid mass of its host galaxy to compute the GWB spectrum; they confirmed the power-law relation and found $A_{\rm{yr}}\sim 10^{-15}$. Subsequently, \citet{Wyithe_Loeb03} employed a comprehensive set of semi-analytical models of dark matter halo mergers and the scaling relation between black hole mass and the velocity dispersion of its host galaxy; they found that the GWB is dominated by sources at redshifts $z\lesssim 2$. Most of recent studies generated consistent results with median values at $A_{\rm{yr}}\approx 1\times 10^{-15}$ \citep{Sesana08GWB,Sesana13GWB,Ravi14GWB}, whereas some suggested that the signal may be a factor of two stronger \citep{Kulier15} or even a factor of four stronger~\citep{McWilliams14}.

Theoretical prediction of the GWB is essential to the interpretation of current PTA upper limits. In \citet{PPTA15Sci}, theoretical models that predict typical signals of $A_{\rm{yr}}\gtrsim 1\times 10^{-15}$ were ruled out with high ($\gtrsim 90\%$) confidence. It was further suggested that the orbital evolution of SMBBHs is either too fast (e.g., accelerated by interaction with ambient stars and/or gas) or stalled. Such a tension between models and observations can be eliminated using different black hole-host scaling relations \citep{SesanaBias16}. Recently, \citet{Middleton2017} quantified this problem within a Bayesian framework by comparing a wide range of models with the PPTA limit and found that only the most optimistic scenarios are disfavoured.

Furthermore, understanding uncertainties in $A_{\rm{yr}}$ predictions is critical to the evaluation of future detection prospects. For example, in \citet{AreWeThere}, the model of \citet{Sesana13GWB} was used in combination with PTA upper limits to calculate the time to detection of the GWB. Based on simple statistical estimates, they suggested $\sim 80\%$ of detection probability within the next ten years for a large and expanding timing array.
Needless to say, this statement hinges on accurate predictions of the minimum GWB.
\citet{nightmare17} attempted to make such a prediction within a semi-analytical galaxy formation model by artificially stalling all SMBBHs at the orbital separation from which GW can drive binaries to merge within a Hubble time; they suggested that a PTA based on the Square Kilometre Array \citep[SKA;][]{Lazio13CQG} is capable of making a detection in this least favourable scenario. Recently, \citet{Bonetti18} and \citet{Ryu18} showed that this scenario might be too pessimistic because triple/multiple interactions can drive a considerable fraction of stalling binaries to merge; both suggested that the GWB is unlikely to be lower than $A_{\rm{yr}}\sim 10^{-16}$.

In this paper, we first assess the implication of current PTA upper limits.
We approach this issue differently from previous studies.
We point out that the local black hole mass function, an electromagnetically determined measure of the number density of black holes as a function of mass, implies a constraint on $A_{\rm{yr}}$. If we suppose all black holes are produced by equal-mass SMBBH mergers, then the black hole mass function gives us a maximum GWB amplitude.


Second, we present a novel Bayesian framework to infer the SMBBH merger rate based on a gold-plated detection of a single system. The derived merger rate can be combined with the system chirp mass to compute the GWB signal amplitude using the practical theorem of \citet{Phinney01}.
We consider several SMBBH candidates with inferred masses and orbital periods, and derive lower bounds on $A_{\rm{yr}}$.



This paper is organized as follows. In Section \ref{AGWBgeneral}, we review the formalism of \citet{Phinney01} and provide two useful equations for quick computation of $A_{\rm{yr}}$. In Section \ref{sec:Ayrmax}, we derive the maximum signal amplitude from several black hole mass functions. In Section \ref{sec:Bayes_Alow} we present the Bayesian framework for inferring the SMBBH merger rate from a single system. We apply this method to several SMBBH candidates to derive plausible lower bounds of $A_{\rm{yr}}$. Finally, Section \ref{sec:sumdis} contains summary and discussions.

\section{The formalism}
\label{AGWBgeneral}
In this section we present a phenomenological model for the GWB formed by a population of SMBBHs in circular orbits. We start from the calculation of $\Omega_{\rm{GW}}(f)$ -- the GW energy density per logarithmic frequency interval at observed frequency $\it{f}$, divided by the critical energy density required to close the Universe today $\rho_{c}=3H_{0}^{2} c^{2}/8\pi G$. Here $H_{0}$ is the Hubble constant. Assuming a homogeneous and isotropic Universe, it is straightforward to compute this dimensionless function as \citep[see][for details]{Phinney01}:
\begin{equation}
\Omega_{\rm{GW}}(f)=\frac{1}{\rho_{c}}\int_{0}^{\infty} \frac{N(z)}{(1+z)} \left.\left(\frac{{\rm{d}} E_{\rm{GW}}}{{\rm{d}}\ln f_{\rm{r}}}\right)\right|_{f_{\rm{r}}=f (1+z)} {\rm{d}}z
\label{omeg},
\end{equation}
where $N(z)$ is the spatial number density of GW events at redshift $z$; the $(1+z)$ factor accounts for the redshifting of GW energy; $f_{\rm{r}}=f (1+z)$ is the GW frequency in the source's cosmic rest frame, and ${\rm{d}}f_{\rm{r}}({\rm{d}}E_{\rm{GW}}/{\rm{d}}f_{\rm{r}})$ is the total energy emitted in GWs within the frequency interval from $f_{\rm{r}}$ to $f_{\rm{r}}+{\rm{d}}f_{\rm{r}}$.

There are two other quantities that are commonly used for the characterization of a GWB, namely, the one-sided spectral density $S_{h}(f)$ and the characteristic amplitude $h_{c}(f)$. They are related to $\Omega_{\rm{GW}}(f)$ by \citep{magg00}:
\begin{equation}
h_{c}^{2}(f)=fS_{h}(f) = \frac{3 H_{0}^{2}}{2 \pi^{2}} f^{-2} \Omega_{\rm{GW}}(f)
\label{hcSh}.
\end{equation}

In the Newtonian limit, the GW energy spectrum for an inspiralling circular binary of component masses $m_1$ and $m_2$ is given by \citep{Thorne87}:
\begin{equation}
\frac{{\rm{d}}E_{\rm{GW}}}{{\rm{d}}f} = \frac{(\pi G)^{2/3} M_c^{5/3}}{3} f^{-1/3} ,
\label{dEdf}
\end{equation}
where $M_c$ is the chirp mass defined as $M_c = M \eta ^{3/5}$, with $M= m_1+m_2$ being the total mass and $\eta = m_1 m_2 /M^{2}$ being the symmetric mass ratio. Equation (\ref{dEdf}) is a good approximation up to the frequency at the last stable orbit during inspiral $f_{\rm{max}} \simeq 4.4\, {\rm{kHz}}/(M/M_{\odot})$. The merger and ringdown processes occur beyond the PTA band and thus their contribution to the GWB is ignored \citep{ZhuBBH11}.

For SMBBHs, if we assume that 1) the binary can reach a separation of $\sim 1$ pc so that dynamical friction becomes ineffective; and 2) the binary hardens through the repeated scattering of stars in the core of the host, then there exists a minimum frequency $f_{\rm{min}}$ for equation (\ref{dEdf}) to be valid as given by \citep{Quinlan96}:
\begin{equation}
f_{\rm{min}} = 2.7\, {\rm{nHz}}\, \left[\frac{m_{1} m_{2}}{(10^{8}\, M_{\odot})^2}\right]^{-0.3}\left(\frac{m_{1}+ m_{2}}{2\times 10^{8}\, M_{\odot}}\right)^{0.2}\, .
\end{equation}

The characteristic amplitude of the GWB formed by a cosmological population of SMBBHs in circular orbits is given by:
\begin{equation}
h_{c}^{2}(f)=\frac{4G^{5/3}}{3c^{2}\pi^{1/3}f^{4/3}} \int\int \frac{{\rm{d}}^{2}N}{{\rm{d}}M_{c} {\rm{d}}z}M_{c}^{5/3} (1+z)^{-1/3}{\rm{d}}M_{c} {\rm{d}}z \, ,
\label{hcCBC1}
\end{equation}
where $\frac{{\rm{d}}^{2}N}{{\rm{d}}M_{c} {\rm{d}}z}$ is the number density (per unit comoving volume) of SMBBH mergers within chirp mass range between $M_{c}$ and $M_{c}+{\rm{d}}M_{c}$, and a redshift range between $z$ and $z+{\rm{d}}z$. The integration is typically performed over $10^{7}-10^{10}\,M_{\odot}$ for $M_{c}$ and $0-2$ for $z$ \citep[see, e.g.,][]{Sesana13GWB}. In this work, we assume that the binary orbital evolution above $f_{\rm{min}}$ is driven solely by GWs. This leads to a $h_{c}\sim f^{-2/3}$ relation for the frequency band (1 nHz $\lesssim f \lesssim$ 100 nHz) of interest to PTAs \citep{Sesana13GWB,Ravi14GWB,McWilliams14}.

To compute $h_{c}(f)$, one needs to know the distribution of SMBBHs in chirp mass and redshift. Previous studies relied on galaxy merger rate as a function of redshift, which are typically derived from cosmological simulations of galaxy formation or observations of galaxy pair fraction combined with galaxy merger timescale, and various black hole-host galaxy scaling relations. In this work, we take a different approach. First of all, we note that equation (\ref{hcCBC1}) can be simplified by assuming that there is no redshift dependency for black hole mass distribution. While this is incorrect, we show in the next sections that its effect is small.

Below we provide two useful equations for computing $h_{c}$ in convenient numerical forms. The first is adapted from \cite{ZhuCBC13} and states
\begin{eqnarray}
h_{c}(f)  &=&  4.8 \times 10^{-16} \left(\frac{10^{4}\times r_0}{{\rm{Mpc}}^{-3}{\rm{Gyr}}^{-1}}\right)^{1/2} \left[\frac{\langle M_{c}^{5/3} \rangle}{(10^{8} M_{\odot})^{5/3}}\right]^{1/2} \nonumber\\&& \times \left(\frac{I_{-4/3}}{0.63}\right)^{1/2} \left(\frac{f}{1{\rm{yr}}^{-1}}\right)^{-2/3}
\label{hcSMBBH2},
\end{eqnarray}
where $r_{0}$ is the local SMBBH merger rate density. A second form is adapted from \citet{Phinney01}
\begin{eqnarray}
h_{c}(f)  &=&  1.5 \times 10^{-16} \left(\frac{10^{4}\times N_{0}}{{\rm{Mpc}}^{-3}}\right)^{1/2} \left[\frac{\langle M_{c}^{5/3} \rangle}{(10^{8} M_{\odot})^{5/3}}\right]^{1/2} \nonumber\\&& \times \left(\frac{I_{-4/3}}{0.86I_{-1}}\right)^{1/2} \left(\frac{f}{1{\rm{yr}}^{-1}}\right)^{-2/3}
\label{hcSMBBH1}.
\end{eqnarray}
where $N_{0}$ is the present-day comoving number density of SMBBH merger remnants. In both equations above, $\langle M_{c}^{5/3} \rangle$ represents the average contribution of coalescing binaries to the energy density of the GWB, and we have defined the following quantity \citep[e.g.,][]{ZhuCBC13}
\begin{equation}
I_{\alpha} = \int_{z_{\rm{min}}}^{z_{\rm{max}}}  \frac{e(z)(1+z)^{\alpha}}{\sqrt{\Omega_{\Lambda}+\Omega_{m}(1+z)^{3}}} {\rm{d}}z
\label{J23CBC}.
\end{equation}
Here $z_{\rm{min}}$ = max(0, $f_{\rm{min}} / f - 1$) and $z_{\rm{max}}$ = min($z_{\star}$, $f_{\rm{max}} / f - 1$) with $z_{\star}$ representing the beginning of source formation, and $e(z)$ is a dimensionless function that accounts for the cosmic evolution of merger rate density. We set $z_{\rm{min}}=0$ and $z_{\rm{max}}=2$ since sources in this redshift range make the majority contribution to the GWB. Throughout this paper, we assume a standard $\Lambda$CDM cosmology with parameters $H_{0}=67.8\, \rm{km}\hspace{0.5mm} \rm{s}^{-1}\hspace{0.5mm} \rm{Mpc}^{-1}$, $\Omega_{m}=0.308$ and $\Omega_{\Lambda}=0.692$ \citep{Planck16cosmo}.

In Equations (\ref{hcSMBBH2}-\ref{hcSMBBH1}), $I_{-4/3}=0.63$ and $I_{-4/3}/I_{-1}=0.86$ are used as fiducial values in the case of $e(z)=1$, i.e., the merger rate remains constant between $z=0$ and $z=2$. Both factors are insensitive to details of $e(z)$. For example, assuming $e(z)=(1+z)^m$, $(I_{-4/3}/0.63)^{1/2}$ only increases from 0.83 for $m=-1$ to 1.26 for $m=1$, whereas the change in $(I_{-4/3}/0.86I_{-1})^{1/2}$ is even smaller -- only varying from 1.01 for $m=-1$ to 0.99 for $m=1$.

The focus of this paper is to the maximum and minimum GWB signal amplitudes, represented by $A_{\rm{yr}}$ for the power-law model
\begin{equation}
h_{c}(f)=A_{\rm{yr}}\left(\frac{f}{1{\rm{yr}}^{-1}}\right)^{-2/3}
\label{hcyr1}.
\end{equation}

\section{The maximum signal}
\label{sec:Ayrmax}
The black hole mass function defines the number density of black holes as a function of mass. Assuming that all the black holes were produced by equal-mass binary mergers and that mass function does not evolve across cosmic time\footnote{The mass function must evolve if all black holes are made from equal-mass mergers. We enforce such an assumption since we are only concerned with the maximum signal.}, the double integral in Equation (\ref{hcCBC1}) can be factorized, leading to the following quantity which defines the maximum signal amplitude allowed by the local black hole mass function
\begin{equation}
\left(A_{\rm{yr}}^{\rm{max}}\right)^{2}=\frac{4G^{5/3}\eta_{\rm{max}}}{3c^{2}\pi^{1/3}f_{\rm{yr}}^{4/3}} \int \frac{{\rm{d}}N}{{\rm{d}}M}(1.1M)^{5/3}{\rm{d}}M \int_{0}^{2}(1+z)^{-1/3} {\rm{d}}z
\label{Ayrmax}\, ,
\end{equation}
where $f_{\rm{yr}}={\rm{yr}}^{-1} \simeq 3.17\times 10^{-8}$ Hz, $\eta_{\rm{max}}=0.25$ is the maximum value of $\eta$ when $m_{1}=m_{2}$, ${\rm{d}}N/{\rm{d}}M$ is the black hole mass function at $z=0$. In Equation (\ref{Ayrmax}) $M$ denotes masses of the final black holes produced by SMBBH mergers. We assume 10\% of rest-mass energy is radiated away in GWs during the inspiral-merger-ringdown process\footnote{Such a radiation efficiency can be seen as an upper bound \citep{Flanagan98}.}, the binary total mass is $1.1M$, which determines the GW emission during inspirals.

\begin{figure}
 \includegraphics[width=0.48\textwidth]{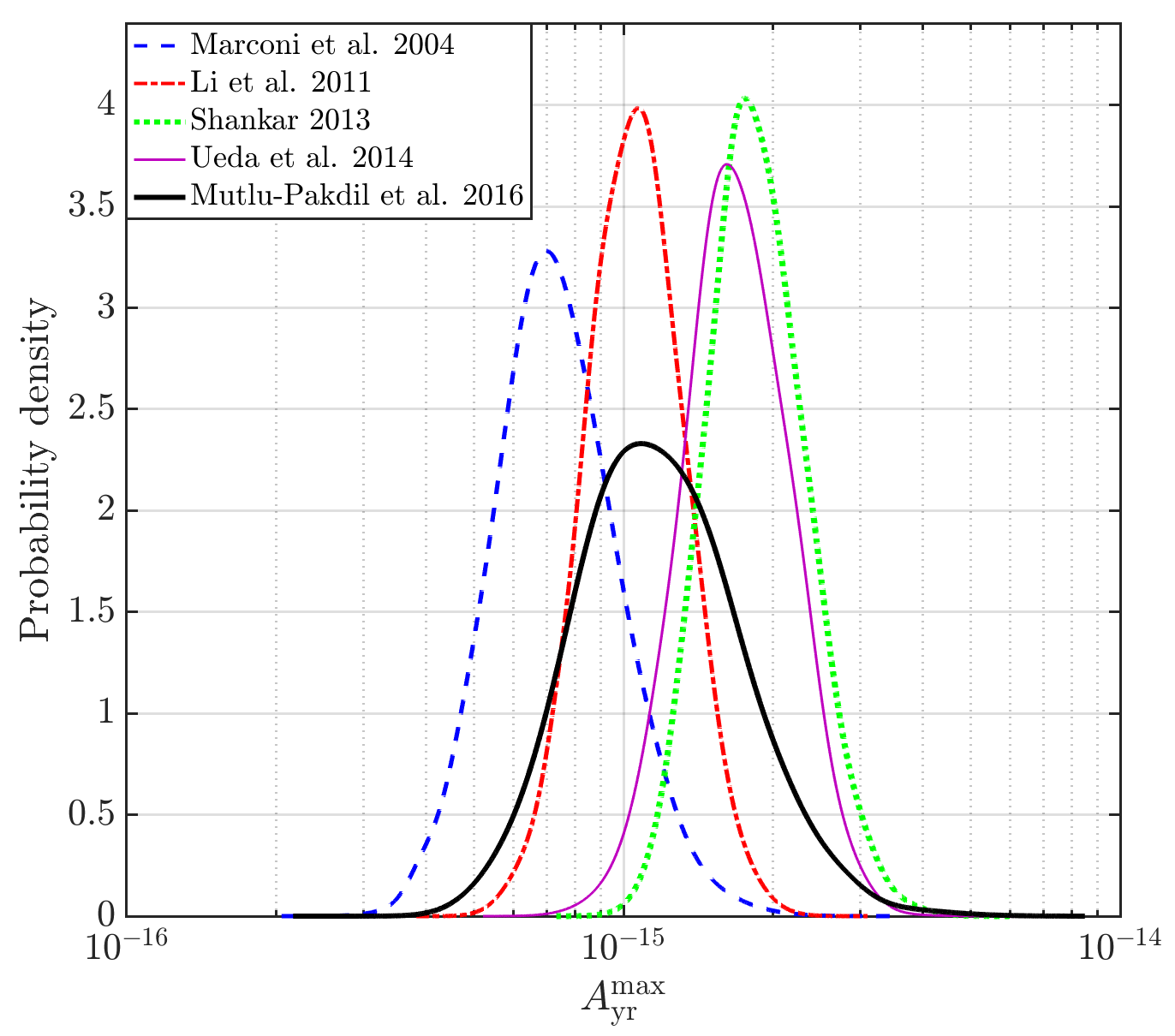}
 \caption{Probability distribution of $A_{\rm{yr}}^{\rm{max}}$ derived from a range of local black hole mass functions (see text for details).}
 \label{fig:AyrUpper}
\end{figure}

Here we consider a range of mass functions determined with different black hole-host scaling relations \citep{Marconi2004,Li2011,Shankar2013,Ueda2014,MP2016}. We provide details of these models in Appendix \ref{appen1-BHMF} and summarize our results here.
In Figure~\ref{fig:AyrUpper}, we plot the probability distribution of $A_\text{yr}^\text{max}$, where the spread is directly translated from measurement uncertainties in black hole mass functions (shown in Figure \ref{fig:BHMF}) through Monte-Carlo simulations.

Among the mass functions considered, the model by \citet{Shankar2013} yields the highest median $A_\text{yr}^\text{max}$ at $1.8\times 10^{-15}$.
While the maximum possible GWB {\em could} be this high, there are good reasons to expect it to be lower.
First, a uniform distribution of mass ratio ($q=m_{2}/m_{1}$) between 0 and 1 gives an average $\langle\eta\rangle=0.193$. This corresponds to a factor of $\sqrt[]{\eta_{\rm{max}}/\langle\eta\rangle}=1.14$. Second, a more realistic radiation efficiency is 5\%, resulting in a factor of $(1.1/1.05)^{5/6}=1.04$.
Third, we make use of the black hole mass function at higher redshifts to compute the following quantity:
\begin{equation}
A_{z}=\left(\frac{\int_{M_{\rm{min}}}^{M_{\rm{max}}} \frac{{\rm{d}}N}{{\rm{d}}M}(z=0)M^{5/3}{\rm{d}}M \int_{0}^{2}(1+z)^{-1/3} {\rm{d}}z}{\int_{M_{\rm{min}}}^{M_{\rm{max}}}\int_{0}^{2} \frac{{\rm{d}}^{2}N}{{\rm{d}}M {\rm{d}}z}M^{5/3} (1+z)^{-1/3}{\rm{d}}M {\rm{d}}z}\right)^{1/2}\, .
\label{eq:Azfactor}
\end{equation}
Using the data presented in \citet{Ueda2014}, we find $A_{z}=1.1$ for $M_{\rm{min}}=10^{7} M_{\odot}$ and $M_{\rm{max}}=10^{10} M_{\odot}$. Overall, a more realistic estimate would be multiplying $A_{\rm{yr}}^{\rm{max}}$ as given by Equation (\ref{Ayrmax}) and illustrated in Figure \ref{fig:AyrUpper} by $\frac{1}{1.14\times 1.04\times 1.10}=0.77$.

To demonstrate the effectiveness of our calculations, we take the black hole mass function used in \citet{Sesana13GWB} (illustrated as solid black lines in the upper panel of their fig. 1), compute $A_{\rm{yr}}^{\rm{max}}$ and then multiply by $0.77$. This gives $5.0\times 10^{-16}< A_{\rm{yr}}<1.1\times 10^{-15}$ with $68\%$ confidence. This matches well with the interval of $3.8\times 10^{-16}< A_{\rm{yr}}<1.1\times 10^{-15}$ at the same confidence level for their fiducial model.

\section{The minimum signal}
\label{sec:Bayes_Alow}
Suppose that there is a gold-plated detection of SMBBH system with measured chirp mass, orbital period and distance.
We show here that it can be used to derive a lower limit on the GWB. A similar approach has been taken to estimate contributions from stellar-mass binary mergers to a GWB signal in the ground-based detector band. For example, the first binary black hole merger GW150914 \citep{GW150914} and the first binary neutron star inspiral GW170817 \citep{GW170817} were both used to infer the corresponding GWB level, for which the uncertainty is dominated by Poissonian errors of the merger rate \citep{GW150914stochastic,GW170817stochastic}.

Here we begin with the description of our framework for performing Bayesian inference on the SMBBH merger rate, and then apply it to several well-established SMBBH candidates.
First, a naive estimator ($\hat{r}_{0}$) for the merger rate density $r_0$ associated with a single SMBBH is
\begin{equation}
\hat{r}_{0}=\frac{1}{\epsilon V_{0} T_{{\rm{c}}}}\, ,
\label{eq:ratelim}
\end{equation}
where $\epsilon$, taking values between 0 and 1, is the detection efficiency whose definition is discussed in detail below; $V_0$ is the comoving volume at the source (comoving) distance $d$, which is given by $(4\pi/3) d^3$;
$T_{{\rm{c}}}$ is the binary coalescence time as a result of gravitational radiation.
For a circular binary, $T_{{\rm{c}}}$ is given by \citep{Thorne87}
\begin{equation}
\label{eq:Tmerge}
T_{\rm{c}}=\frac{5c^5}{256(2\pi f_{0})^{8/3}(G M_{c})^{5/3}}\, ,
\end{equation}
where $f_{0}=1/P_{\rm{b}}$ is the observed binary orbital frequency with $P_{\rm{b}}$ being the orbital period. For a binary with measured eccentricity $e_{0}$, the coalescence time can be well approximated by multiplying the above equation by $\exp({-4e_{0}^2})$ if $e_{0}\lesssim 0.7$.

The electromagnetic detection of an SMBBH will encode two pieces of information: that there is at least $N=1$ SMBBH and that it was observed at a (co-moving) distance of $\hat{d}$.
The true distance is $d$ (no hat).
The detection efficiency $\epsilon$ in Equation (\ref{eq:ratelim}) is used to account for two factors. First, it accounts for the incompleteness of the survey that discovered the SMBBH system. For example, the search may have only covered part of the sky. Second, it accounts for the fact that nearby binaries could be missed even if they are in the sky region included in the survey. For example, if there are two binaries with identical component masses, distances, and orbital period, but with different viewing angles, one may be detectable while the other is not \citep[see, e.g., the scenario discussed in][]{Orazio15}. For the purpose of estimating the minimum signal, we assume $\epsilon =1$, which underestimates the merger rate.

In reality, many SMBBH candidates, such as those considered in our study, have been discovered serendipitously; they were not discovered via a systematic survey. However, we can model these serendipitous discoveries by parameterizing the unknown detection efficiency. We parameterize this efficiency curve as a step function, which is unity for $d<d_\text{max}$ and zero for $d>d_\text{max}$. This is a reasonable approximation given the rate at which distant objects become dimmer with distance. Having made this assumption, we use the data itself to infer $d_\text{max}$. The physical interpretation of dmax is that it is the ``effective maximum detection distance" for whatever measurements led to the discovery of an SMBBH candidate.
We model the likelihood of the observation of an SMBBH as ${\cal L}(N,\widehat{d}|d,r_0, d_\text{max})$ where $N=1$ is the number of observed SMBBH in some observable volume. The likelihood is a Poisson distribution
\begin{align}
{\cal L}(N,\hat{d}|d, r_0, d_\text{max};\epsilon=1) = &
\lambda e^{-\lambda} \delta(d-\hat{d})
\end{align}
where $\lambda$, the average number of SMBBH,
\begin{align}
\lambda = & r_0 V T_c \\
= & r_0 \left(\frac{4}{3}\pi \,
d_\text{max}^3\right) T_c ,
\end{align}
depends on the rate $r_0$, the visible volume $V=(4\pi/3) d^3_{\text{max}}$, and the binary coalescence time $T_c$.
By introducing a delta function $\delta(d-\hat{d})$, we assume that the distance is measured with high precision.

Assuming that sources are uniformly distributed in comoving volume up to $d_\text{max}$, the conditional probability of observing one source at distance $d$ is given by
\begin{align}
\pi(d|d_\text{max}) = \frac{3d^2}{d_\text{max}^3} \Theta(d_\text{max}-d) \, ,
\end{align}
where $\Theta$ is the Heaviside step function: $\Theta(x)=1$ for $x\geq 0$ and $\Theta(x)=0$ otherwise.
Applying Bayes' theorem, we obtain a posterior distribution for the rate $r_0$ and the maximum distance $d_{\rm{max}}$:
\begin{align}
p(r_0,d_\text{max},d|N,\hat{d}) \propto &
{\cal L}(N,\hat{d}|d,r_0,d_\text{max}) \, \pi(d|d_\text{max})
\, \pi(r_0) \, \pi(d_\text{max}) .
\end{align}
Marginalizing over $d$, we obtain
\begin{align}
p(r_0,d_\text{max}|N,\hat{d}) \propto &
\bigg(\lambda e^{-\lambda}\bigg) 
\bigg[\frac{3\hat{d}^2}{d_\text{max}^3} \Theta(d_\text{max}-\hat{d})\bigg]
\pi(r_0)\pi(d_\text{max})
.
\end{align}
Marginalizing over $d_\text{max}$ lead to a posterior on $r_0$. 
We assume a log-uniform prior for $r_0$ and a uniform prior for $d_\text{max}$. Comparing a log-uniform prior with uniform prior, our choice of priors is conservative as it put more a priori weights on lower $r_0$ and larger $d_\text{max}$. The prior range for $r_0$ and $d_\text{max}$ is from 0 to infinity and from $\hat{d}$ to infinity. 

We convert the posterior on $r_0$ to a posterior on $A_\text{yr}$ using Equation (\ref{hcSMBBH2}) by setting $I_{-4/3}=0.63$ and $\langle M_{c}^{5/3} \rangle=\bar{M}_{c}^{5/3}$. Here $\bar{M}_{c}$ is the measured chirp mass for the SMBBH system in question; such a treatment justifies the assumption in Equation (\ref{hcSMBBH2}) that the mass distribution is independent of redshift.
As already mentioned, $I_{-4/3}$ is insensitive to the merger rate evolution within $z\leq 2$; the associated uncertainty is $\approx 50\%$, which is much smaller than the Poissonian uncertainty of the merger rate.

We apply the framework developed here to some well-established individual SMBBH candidates with reported estimates of masses, orbital period and eccentricity. Table \ref{tab:smbbhs} lists key parameters of these candidates and median values of $r_0$ and $A_{\rm{yr}}$. Note that all five binary candidates considered here are in the PTA band, i.e., with $\mathcal{O}(10)$-yr periods or sub-pc orbital separations. SMBBH candidates with much longer periods contribute negligibly to the conservative merger rate estimates derived here ($r_{0}\propto T_{\rm{c}}^{-1}\propto P_{b}^{-8/3}$).

\begin{figure}
 \includegraphics[width=0.48\textwidth]{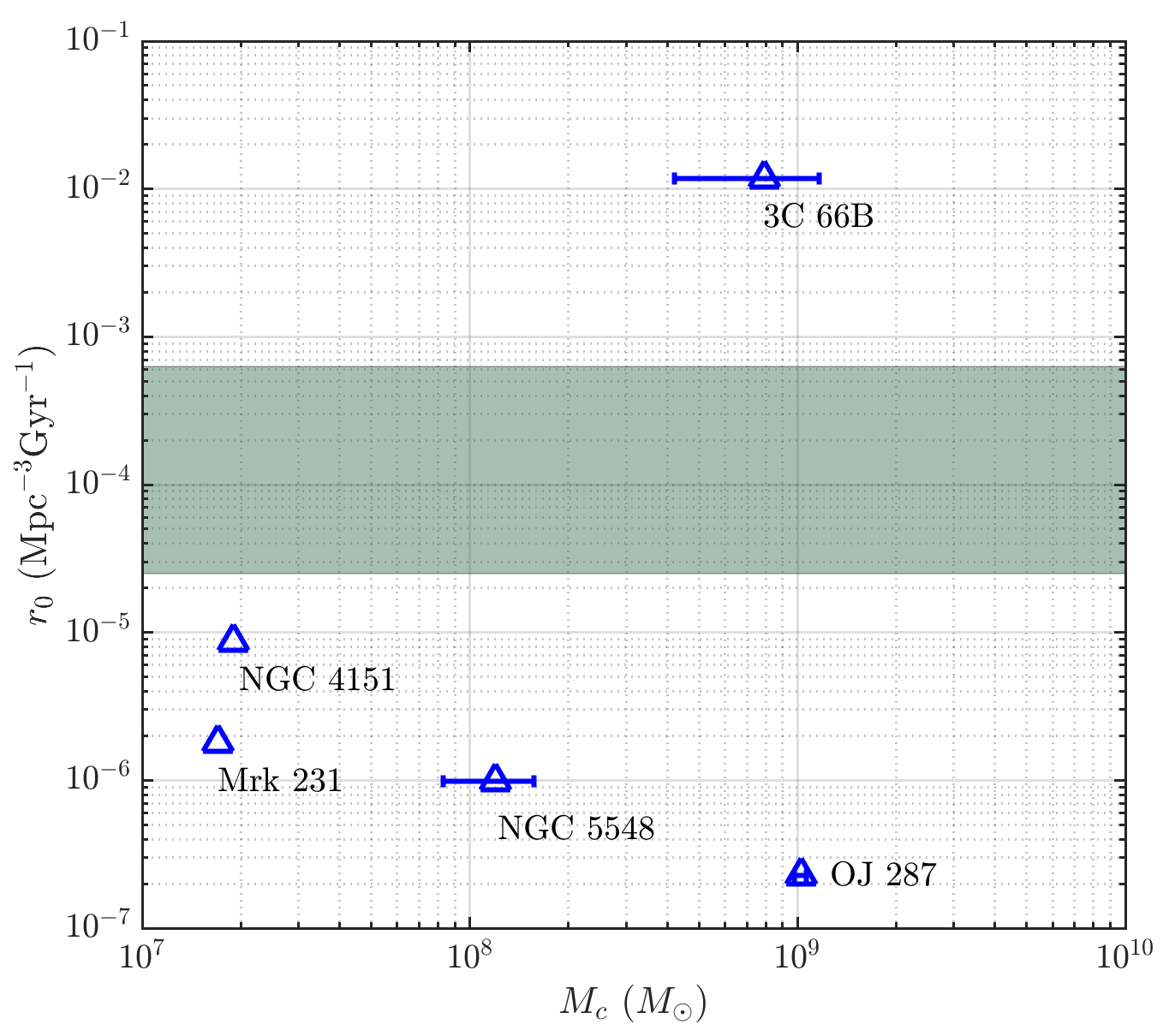}
 \caption{The local SMBBH merger rate density $r_0$ as a function of the binary chirp mass ($M_c$) inferred from several binary candidates; Blue triangles mark 68\% confidence lower limits of $r_0$ (see Table \ref{tab:smbbhs} for details). The shaded horizontal band corresponds to estimates of galaxy merger rate presented in \citet{Conselice14}. To ensure the most conservative estimate, it was assumed the search that discovered each object had a completeness limit out to $d_\text{max}$ (see text for details).}
 \label{fig:r0Mc}
\end{figure}

In Figure \ref{fig:r0Mc} we plot the inferred $r_0$ as a function of chirp mass.
There are a couple of features worthy of remark. First, it is apparent that the derived merger rate for 3C 66B is four orders of magnitude above that of OJ 287 or NGC 5548, with both having similar chirp masses. 
The reason that 3C 66B produces such a high merger rate is due to its very short merger time. If it is a true binary, it is expected to merge in only 500 years, whereas others will typically merge in $\gtrsim$1 Myr. OJ 287 is an exception to the previous sentence: if it is real, it will merge in $10^4$ years. However, it is also much further away than 3C66B, which prevents the merger rate from being as high.
The high rate implied by 3C 66B cannot be explained by the uncertainty in mass estimates. Reducing the chirp mass by a factor of two, the typical uncertainty claimed in \citet{3C66B}, would only decrease merger rate by a factor of $2^{5/3}=3.2$. Furthermore, if 3C 66B is a true SMBBH system, the implied GWB amplitude is nearly two orders of magnitude above the current PTA upper limit $A_{\rm{yr}}\leq 1\times 10^{-15}$ (see Table \ref{tab:smbbhs}). In Section \ref{sec:3c66b}, we use the PTA limit to rule out parameter space in $(m_{1},\, m_{2})$ for 3C 66B.

Second, $r_0$ inferred from the other four systems are consistent with current estimates of galaxy merger rate density $r_g$, which is shown as a shaded horizontal band in Figure \ref{fig:r0Mc}. For the purpose of illustration, the lower edge of this band corresponds to the lowest estimate ($r_{g}\sim 2.5\times 10^{-5}\,{\rm{Mpc}}^{-3}{\rm{Gyr}}^{-1}$) presented in \citet{Conselice14} for galaxy stellar mass above $10^{11} M_{\sun}$ within $z\leq 1$, and the upper edge corresponds to the highest estimate ($r_{g}\sim 6.3\times 10^{-4}\,{\rm{Mpc}}^{-3}{\rm{Gyr}}^{-1}$) for galaxy stellar mass above $10^{10} M_{\sun}$ within $z\leq 1$; see bottom panels of fig. 13 therein.

In the following two subsections, we discuss two special candidates: 3C 66B and OJ 287.

\begin{table*}
 \begin{tabular}{lccccccccccc}
  \hline
  \multirow{2}{*}{Name} & \multirow{2}{*}{$z$} & $d_{L}$ & $m_1$ & $m_2$ & $P_{\rm{b}}$ & \multirow{2}{*}{$e_0$} & \multirow{2}{*}{Ref.} & $M_c$ & $T_{\rm{c}}$ & $r_0$ & $A_{\rm{yr}}$\\
  &  & (Mpc)& ($10^{8} M_{\odot}$) & ($10^{8} M_{\odot}$) &(yr) & & & ($10^{8} M_{\odot}$) & (Myr) & (${\rm{Mpc}}^{-3}{\rm{Gyr}}^{-1}$) &  ($10^{-16}$)\\
  \hline
  3C 66B & 0.0213 & 95.7 & 12$^{+5}_{-2}$ & 7.0$^{+4.7}_{-6.4}$ & 1.05 & 0 & (1) & 7.92$\pm3.7$  & $5.1\times 10^{-4}$ & 0.1 &  860  \\
   OJ 287 & 0.3056 & 1635 & 183$\pm1$ & 1.5$\pm 0.1$ & 12.1 & 0.7 & (2) & 10.23$\pm 0.43$ & $1.1\times 10^{-2}$ & $2.0\times 10^{-6}$ & 4.7  \\
  NGC 5548 & 0.0172 & 77.1 & 1.51$\pm0.48$ & 1.26$\pm0.4$ & 14.1 & 0.13 & (3) & 1.20$\pm0.37$  & 11.5 & $8.6\times 10^{-6}$ & 1.6  \\
  NGC 4151 & 0.0033 & 14.6 & 0.44 & 0.12 & 15.9 & 0.42 & (4) & 0.19 & 183.5 & $7.5\times 10^{-5}$ & 1.1 \\
  Mrk 231 & 0.0422 & 153 & 1.46 & 0.04 & 1.2 & 0 & (5)  & 0.17 & 0.43 &  $1.6\times 10^{-5}$ & 0.4  \\
  \hline
 \end{tabular}
 \caption{Parameters of SMBBH candidates, including redshift ($z$), luminosity distance ($d_L$), black hole masses ($m_1$ and $m_2$), the observed binary orbital period ($P_b$) and eccentricity ($e_0$). Note that $e_{0}=0$ was assumed for 3C 66B and Mrk 231 in the original publications. Mass errors for NGC 4151 and Mrk 231 are either unavailable or difficult to interpret and thus ignored. References: (1). \citet{3C66B}, (2). \citet{OJ287spin16}, (3). \citet{NGC5548Li16}, (4). \citet{NGC4151Bon12}, (5). \citet{Mrk231}. We also derive the binary chirp mass ($M_c$), the coalescence time $T_{\rm{c}}$ and the median estimates of the SMBBH merger rate $r_0$ and the GWB signal amplitude $A_{\rm{yr}}$.}
  \label{tab:smbbhs}
\end{table*}

\subsection{3C 66B}
\label{sec:3c66b}
The elliptical galaxy 3C~66B is located at a redshift of 0.0213. 
\citet{3C66B03} observed variations in the radio core position with a period of 1.05 years and interpreted this as due to the orbital motion of an SMBBH. The proposed binary system, with inferred total mass of $5.4\times 10^{10} M_{\sun}$ and mass ratio of 0.1, was subsequently ruled out with 95\% confidence by \citet{Jenet04} using timing observations of PSR B1855+09 presented in \citet{Kaspi94}. \citet{3C66B} performed follow-up observations of the source and obtained significantly lower mass estimates -- $m_{1}=12_{-2}^{+5}\times 10^{8} M_{\sun}$ and $m_{2}=7.0_{-6.4}^{+4.7}\times 10^{8} M_{\sun}$ assuming a circular orbit. We note that this is below current PTA sensitivities on individual SMBBHs \citep[see, e.g.,][]{ZhuPPTACW14}.

\begin{figure}
 \includegraphics[width=0.48\textwidth]{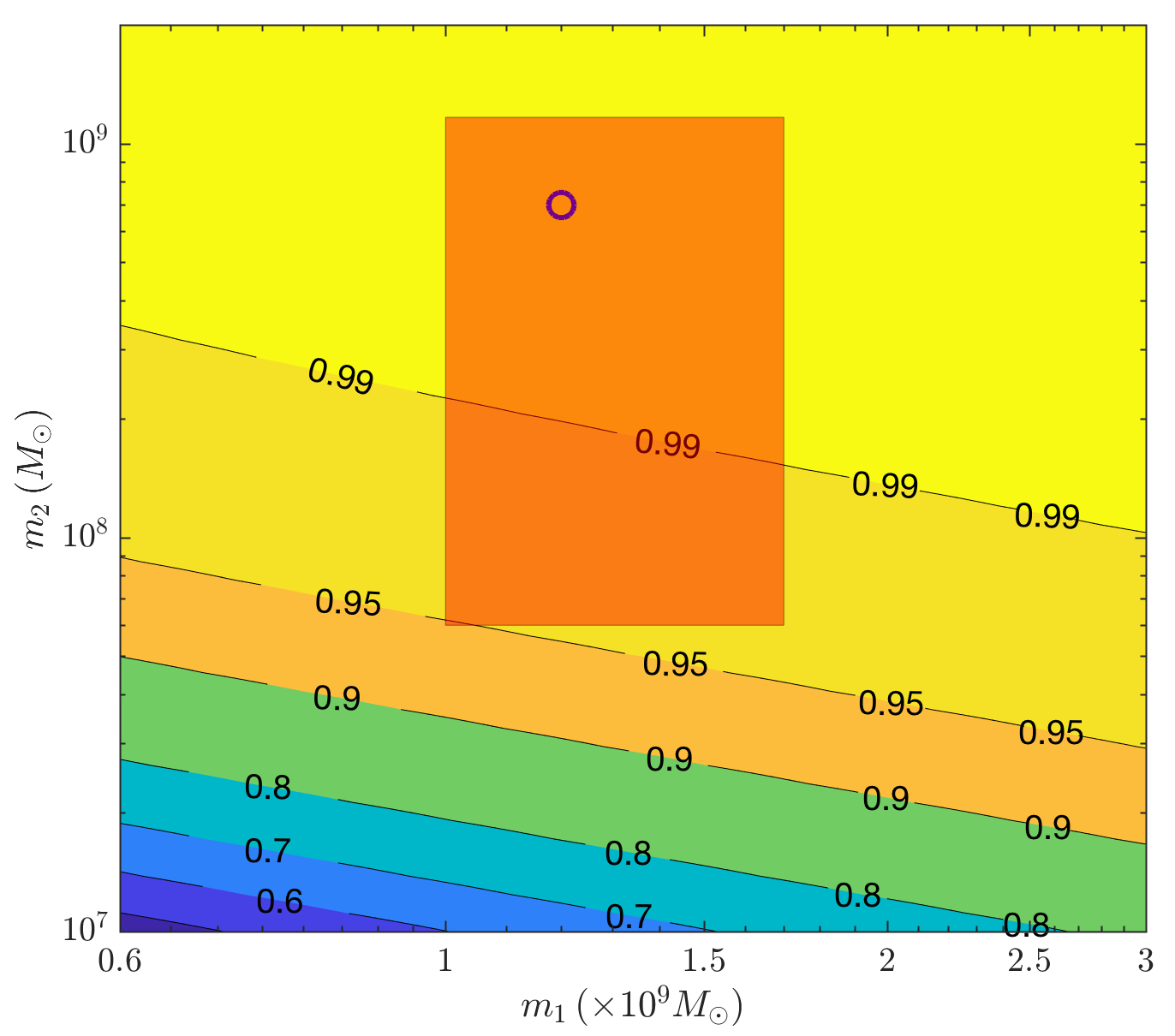}
 \caption{The probability $p(A_{\rm{yr}}>10^{-15})$ if 3C 66B is a true SMBBH system. The red shaded box encompasses the (presumably) 68\% confidence intervals reported in \citet{3C66B}.}
 \label{fig:3c66bm1m2}
\end{figure}

Following the procedure described above, we compute the probability distribution of the GWB signal amplitude $A_{\rm{yr}}$ if we take 3C 66B as a true SMBBH system. We fix the orbital period at 1.05 years; Given its small uncertainty of 0.03 years, our results are not significantly affected by such a simplification.
Figure \ref{fig:3c66bm1m2} shows the probability that $A_{\rm{yr}}>10^{-15}$ for a range of masses\footnote{Note that stronger statements can be made if the full posterior of $A_{\rm{yr}}$ from PTAs is used to perform the consistency test between a model and the data \citep{PPTA15Sci}.}. The red shaded box encompasses the (presumably 68\% confidence) error region reported in \citet{3C66B}, whereas the blue circle marks the median estimate. One can see that the median masses can already be ruled out by current PTA upper limits with more than 99\% confidence, whereas the entire error box is in tension with PTA observations with 95\% probability. This implies that 3C 66B is unlikely to contain an SMBBH.

\subsection{OJ 287}
\label{sec:oj287}
OJ 287 is a BL Lac object with 12-year quasi-periodic variations in optical light curves. Its observations dated back to 1890s and it was proposed as an SMBBH candidate first by \citet{OJ287_88} with later refinement by \cite{OJ287Nature08}. Here the model is that a secondary black hole is in an eccentric orbit around a primary black hole, crossing the accretion disk of the primary once every 12 years. This binary system is described with the following parameters: $(m_{1},m_{2})=(1.83\times 10^{10}, 1.5\times 10^{8})\, M_{\odot}$, orbital eccentricity $e=0.7$, observed orbital period $P_{\rm{b}}=12.07$ yr and redshift $z=0.3056$ \citep{OJ287spin16}, leading to a short merger time $T_{\rm{c}}=1.6 \times 10^{4}$ yr. The spin of the primary black hole is ignored as its effect on the GWB at low frequencies is negligible \citep{ZhuBBH11}.

\begin{figure}
 \includegraphics[width=0.48\textwidth]{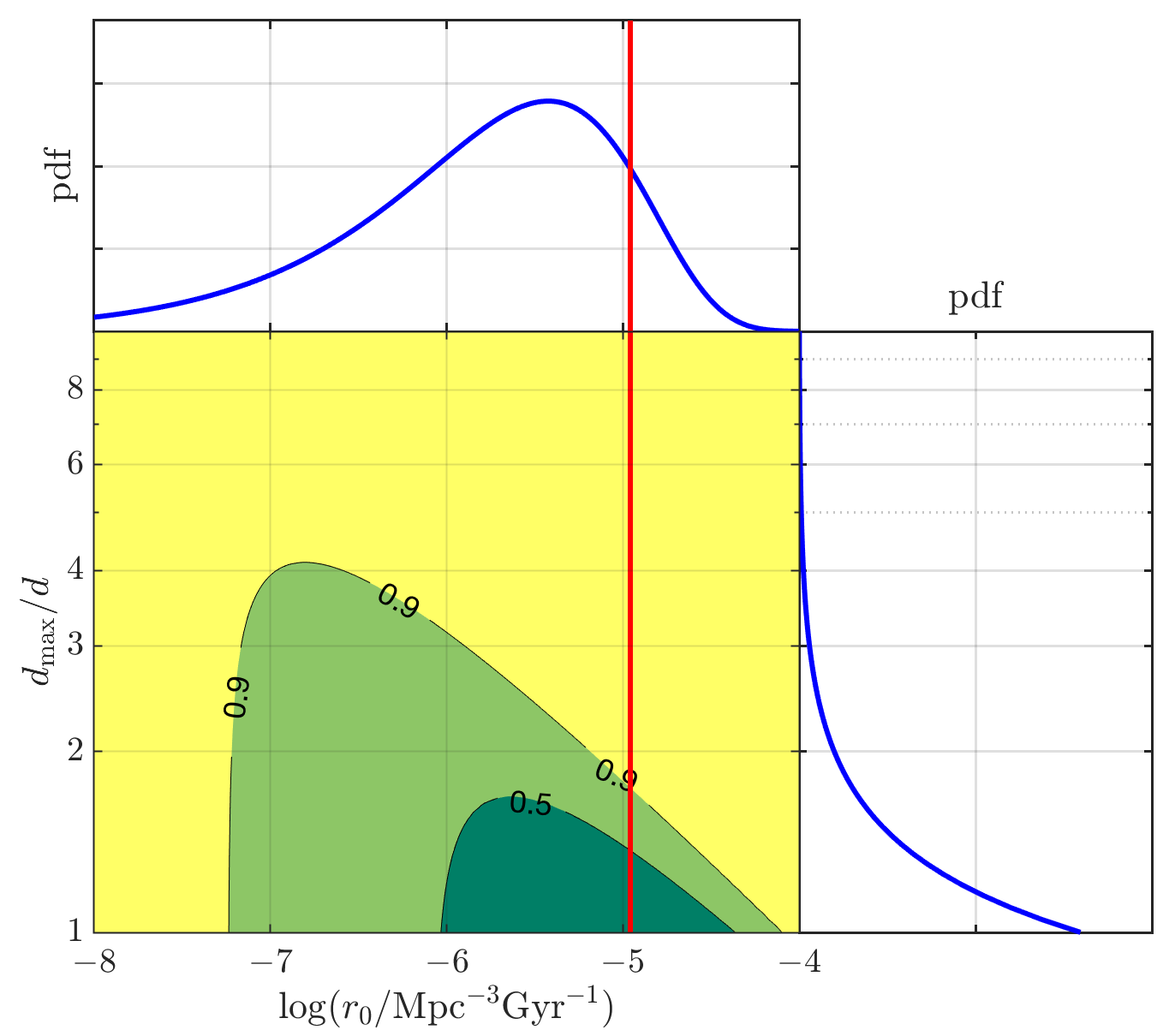}
 \caption{The posterior probability density on the merger rate density $r_0$ of OJ 287-like SMBBHs and the horizon distance $d_{\rm{max}}$. The red vertical line marks the naive rate estimator $\hat{r}_{0}$ given in Equation (\ref{eq:ratelim}). The contour lines mark the confidence regions.}
 \label{fig:r0dmax}
\end{figure}

We compute probability distribution $p(A_{\rm{yr}})$ under the assumption that OJ 287 is a true SMBBH system.
First, the naive estimator of merger rate given by Equation (\ref{eq:ratelim}) is $1.1\times 10^{-5}\,{\rm{Mpc}}^{-3}{\rm{Gyr}}^{-1}$. Figure \ref{fig:r0dmax} shows the posterior probability of the merger rate density $r_0$ and $d_{\rm{max}}$. The red vertical line marks the naive estimator $\hat{r}_{0}$. After marginalizing over the unknown $d_{\rm{max}}$, we find the merger rate density of OJ 287-like SMBBHs to be in $(2.3 \times 10^{-7}, 8.5\times 10^{-6})\,{\rm{Mpc}}^{-3}{\rm{Gyr}}^{-1}$ with 68\% confidence.

\begin{figure}
 \includegraphics[width=0.48\textwidth]{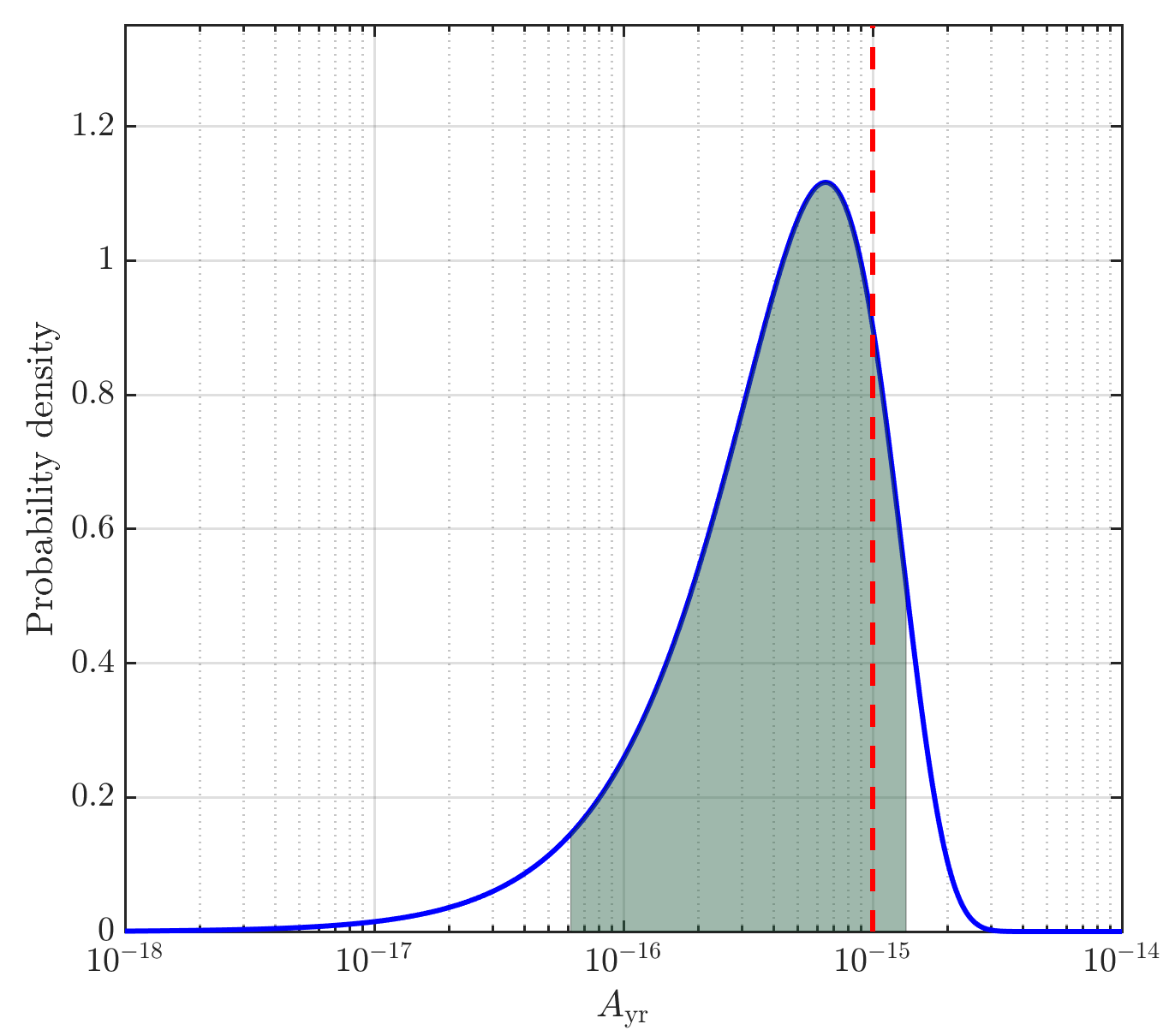}
 \caption{The probability distribution of the GWB amplitude $A_{\rm{yr}}$ under the assumption that OJ 287 is a true SMBBH system. The red vertical line marks the PTA upper limit of $10^{-15}$. The shaded region corresponds to 90\% confidence interval.}
 \label{fig:pAyr_OJ287}
\end{figure}

Figure \ref{fig:pAyr_OJ287} shows the probability distribution of the GWB amplitude $A_{\rm{yr}}$ transformed from the marginalized posterior distribution of merger rate. We find that 1) $A_{\rm{yr}}$ lies between $1.6\times 10^{-16}$ and $9.7\times 10^{-16}$ with 68\% confidence and 2) $A_{\rm{yr}}>6.1\times 10^{-17}$ with 95\% confidence.

\section{Summary and Discussions}
\label{sec:sumdis}

\begin{figure}
 \includegraphics[width=0.48\textwidth]{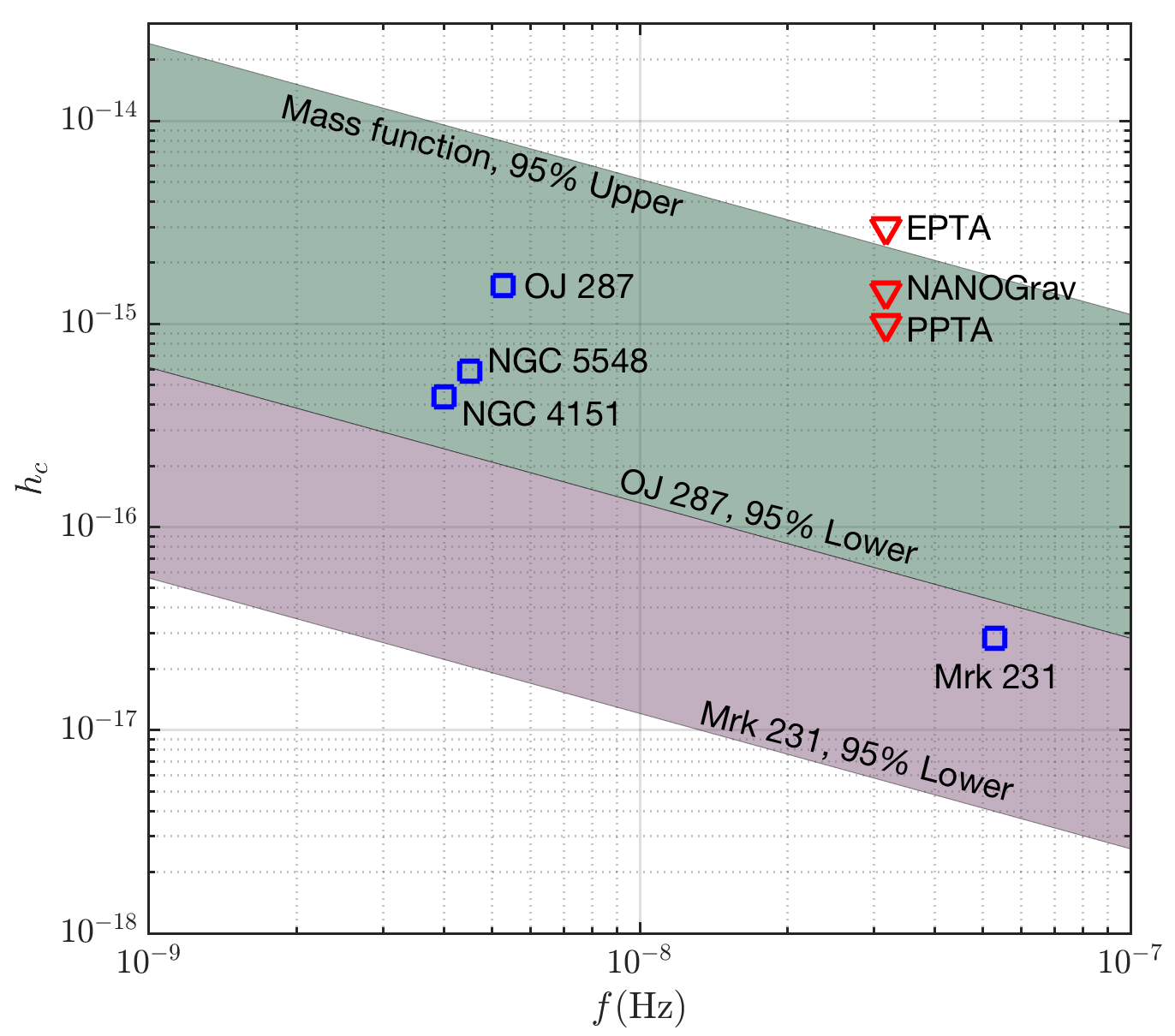}
 \caption{Characteristic amplitude ($h_{c}$) of the GWB signal from SMBBHs. The shaded bands are determined by the minimum and maximum amplitudes (all at 95\% confidence level) derived in this work. Red triangles are 95\% confidence upper limits from various PTA experiments. Blue squares are the median estimates of $A_{\rm{yr}}$ listed in Table \ref{tab:smbbhs} extrapolated to twice the observed orbital frequency for several SMBBH candidates.}
 \label{fig:hcf_bound}
\end{figure}

We summarize our main results in Figure \ref{fig:hcf_bound}.
The shaded regions are determined by the minimum and maximum amplitudes derived in this work.
The maximum is the 95\% confidence upper limit $A_{\text{yr}} \leq 2.4\times 10^{-15}$, if we consider five models of black hole mass function to be equally likely in Section \ref{sec:Ayrmax}.
The  minimum is the 95\% confidence lower limit $A_{\text{yr}} \geq 6.1\times 10^{-17}$ if OJ 287 is a true SMBBH system.
If at least one of OJ 287, NGC 5548, NGC 4151 and Mrk 231 host a true SMBBH with parameters inferred in the literature, $A_{\text{yr}}\geq 5.6\times 10^{-18}$ with 95\% confidence.
Red triangles mark the 95\% confidence upper limits from three PTA experiments\footnote{Note that such PTA upper limits were set for a $-2/3$ power-law GWB rather than at a single frequency of 1 yr$^{-1}$. In fact, because of the steep spectrum of the GWB signal, PTA experiments are most sensitive to a much lower frequency. This frequency is comparable to $1/T_{\rm{obs}}$ with $T_{\rm{obs}}\gtrsim 10$ yrs being the data span.}: $1\times 10^{-15}$ from PPTA \citep{PPTA15Sci}, $1.45\times 10^{-15}$ from NANOGrav \citep{NANOGrav11GWBlimit} and $3\times 10^{-15}$ from EPTA \citep{EPTA15GWBlimit}.

Our calculations of the maximum amplitude provide a straightforward interpretation of PTA upper limits -- if the GWB signal was stronger, we would have been able to see more single black holes left over from SMBBH merger events. We conclude that existing PTA limits constrain only the extremely optimistic models, in agreement with the recent work by \citet{Middleton2017}.

While current PTAs steadily increase their sensitivities and next-generation PTAs are being commissioned or planned for new telescopes such as FAST \citep{FAST11}, MeerKAT \citep{MeerTime} and ultimately the SKA, it is critical to understand what is the minimum level of GWB from the cosmic population of SMBBHs.

We presented in this paper a novel Bayesian framework to estimate the minimum amplitude of this highly-sought signal. We demonstrated that a single gold-plated detection of an SMBBH system in the local Universe immediately implies a lower limit on the GWB.
We applied our framework to several well-established sub-parsec SMBBH candidates. We found that 3C 66B is unlikely to host an SMBBH system because if it was, it would 1) suggest a GWB signal that is inconsistent with existing upper limits at high ($>99\%$) confidence and 2) indicate an SMBBH merger rate that is two orders of magnitude higher than current estimates of galaxy merger rate.

If OJ 287 is a true SMBBH system with parameters suggested in \citet{OJ287spin16}, a median GWB is to have $A_{\rm{yr}}= 4.7\times 10^{-16}$. While this may sound like a good news for PTAs, we note, however, that a lower total mass of $4\times 10^{8}\,M_{\sun}$ was derived in \citet{OJ287LiuWu} and further supported recently by \citet{OJ287lowmass17}, in contrast to $1.8\times 10^{10}\,M_{\sun}$ used in our calculations. This would reduce the $A_{\rm{yr}}$ estimate by a factor of 600 (since $A_{\rm{yr}}\propto M_{c}^{5/3}$) if other parameters remain unchanged.

Blue squares in Figure \ref{fig:hcf_bound} show the median predictions of the GWB amplitude at twice the orbital frequency for several SMBBH candidates. These sources are all in the GW dominant regime. Apart from being interesting targets for continuous GW searches, they collectively suggest a sizeable GWB signal for PTAs. Without looking into specific details of each source for which no quantified statistical significance is available, a simple argument is that $A_{\rm{yr}}>5.6\times 10^{-18}$ at 95\% confidence if at least one of these candidates is a true binary black hole with parameters inferred in the literature.

We suggest that advances in the following areas will be helpful to improve our predictions. First, quantified statistical significance of the SMBBH candidates can be built into our framework to produce more robust GWB predictions. Second, better understanding of the discovery efficiency, sensitive volume and survey completeness of various observational campaigns that search for sub-parsec SMBBHs will lead to tighter constraints on the SMBBH merger rate and the GWB amplitude.

Finally, our calculations focused on the $h_{c}\sim f^{-2/3}$ power spectrum. The actual signal spectral shape is likely to deviate from this. First, the small number of binaries contributing to the background reduces signal power at $f\gtrsim 1\,{\rm{yr}}^{-1}$ \citep{Sesana08GWB,Ravi2012}. Second, effects of binary eccentricity \citep{Enoki07,HuertaEcc15} and the interaction between SMBBHs and their environments \citep{Sesana04,Ravi14GWB} are known to attenuate the signal power at $f\lesssim 0.1\,{\rm{yr}}^{-1}$ \citep[see][for details]{Kelley2017b}. Nevertheless, the method presented here\footnote{Our codes are publicly available at \url{https://github.com/ZhuXJ1/PTA_GWBminmax}} is especially useful for obtaining leading-order predictions for the GWB signal. In particular, when a new SMBBH candidate is discovered, our method allows quick evaluation of its implications for the GWB, and potentially enable constraints to be placed on black hole masses.
In short, an {\em unambiguous} SMBBH detection will have immediate implications to PTAs.

\section*{Acknowledgements}
We thank the referee for very useful comments. We also thank Alberto Sesana, Yuri Levin, Wang Jian-Min, Li Yan-Rong and Lu Youjun for insightful discussions. X.Z. \& E.T. are supported by ARC CE170100004.
E.T. is supported through ARC FT150100281.
W.C. is supported by the {\it Ministerio de Econom\'ia y Competitividad} and the {\it Fondo Europeo de Desarrollo Regional} (MINECO/FEDER, UE) in Spain through grant AYA2015-63810-P.




\bibliographystyle{mnras}
\bibliography{Ref}




\appendix

\section{The local black hole mass function}
\label{appen1-BHMF}

\begin{figure}
 \includegraphics[width=0.48\textwidth]{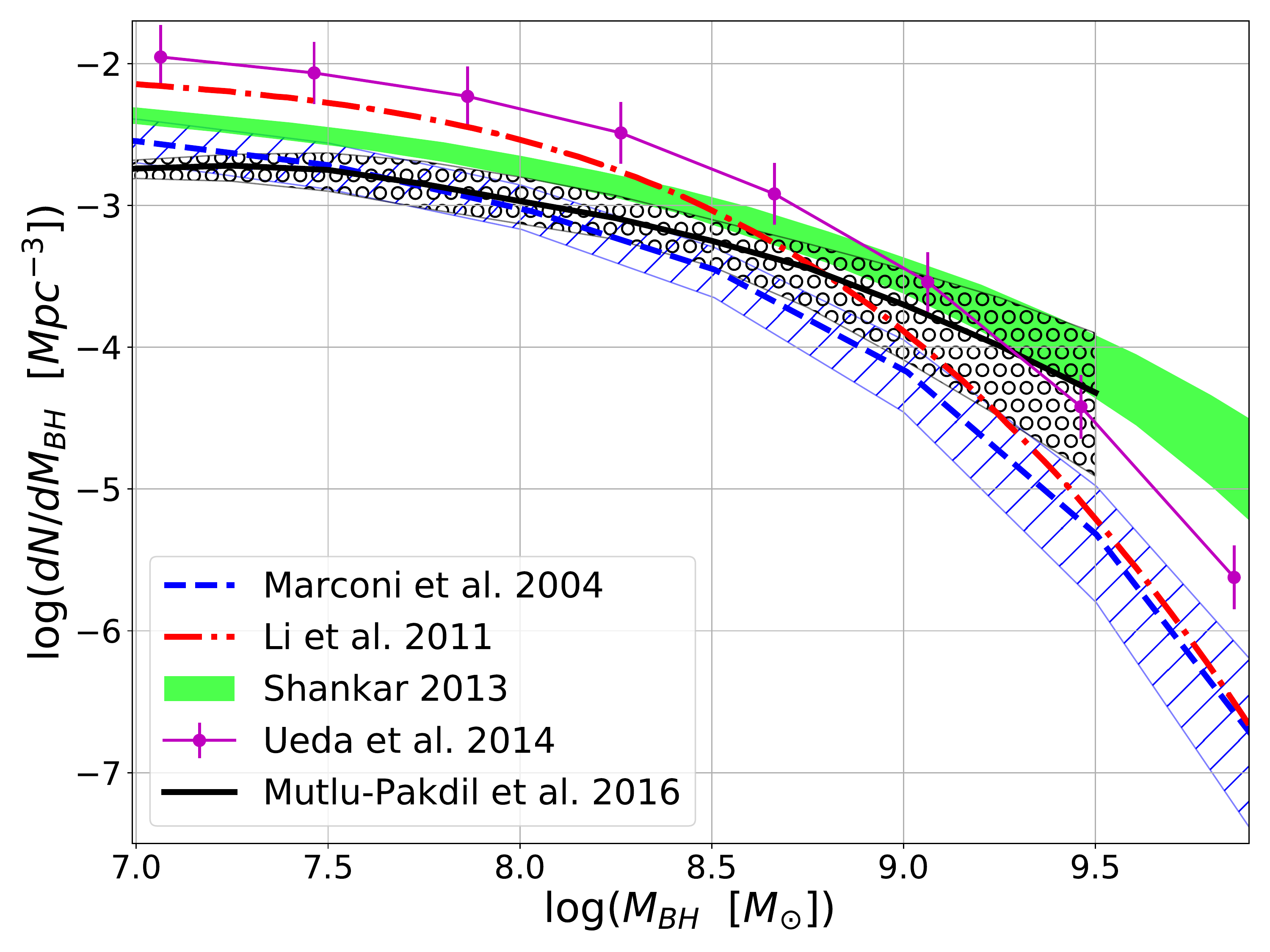}
 \caption{Five models of the local black hole mass function used in Section \ref{sec:Ayrmax} and for upper bounds on the GWB amplitude illustrated in Figure \ref{fig:AyrUpper}. The 1-$\sigma$ uncertainties are shown as filled regions, shaded areas or error bars. As no uncertainty was provided in the model of \citet{Li2011}, we assume 20\% variation for  $\log ({\rm{d}}N/{\rm{d}}M_{BH})$.}
 \label{fig:BHMF}
\end{figure}

Black hole masses ($M_{BH}$) are normally estimated from the black hole-host galaxy scaling relations. A comparison among the local black hole mass functions based on different scaling relations can be found in, for example, fig. 1 of \citet{yu2008} and fig. 5 of \citet{Shankar2009}. In this work, we consider five models that are derived using different methods. They are visually represented in Fig.~\ref{fig:BHMF} and we provide brief descriptions below.

\citet{Marconi2004} adopted both the $M_{BH} - \sigma$ and the $M_{BH} - L_{Bulge}$ relations and found similar results. Here we use their model from all types of galaxies. 
The model of \citet{Li2011} is obtained using the galaxy catalogue of the UKIDSS Ultra Deep Survey combined with the empirical correlation between $M_{BH}$ and spheroid mass (the $M_{BH}-M_{sph}$ relation).
We also include the mass function from \citet{Shankar2013} and take the model that assumes all local galaxies follow the early-type $M_{bh}-\sigma$ relation of \citet{McConnell2013}. This model is suggested to be an {\it upper limit} to the local mass function.

Recently, \citet{MP2016} estimated black hole masses with the $M_{BH} - P$ relation (with $P$ being the galactic spiral arm pitch angle) for late-type galaxies and the $M_{BH} - n$ relation (with $n$ being the S\`ersic index) for early-type galaxies. This model gives the lowest value at the lower mass end. The four models mentioned so far are all based on optical observations. We further consider the model by \citet{Ueda2014} in which the mass function was derived from the X-ray luminosity function of active galactic nuclei. In this case, the X-ray luminosity can be related to the mass accretion rate onto black holes; the mass function can then be derived through the continuity equation.

As one can see from Fig.~\ref{fig:BHMF}, the five models are broadly consistent with each other. Overall, the uncertain range is about an order of magnitude between $10^7$ and $10^{9}\, M_{\sun}$ and larger at the higher mass end.
We integrate these mass functions from $10^7$ up to $10^{10} M_{\sun}$ to compute the number density $N_0$ and the mass density $M_0$:
\begin{equation}
N_0 = \int_{10^{7} M_{\sun}} \Phi(M) {\rm{d}}M\,,\, \hspace{5mm}M_0 = \int_{10^{7} M_{\sun}} M\Phi(m) {\rm{d}}M\, .
\label{eq:N0M0}
\end{equation}
Here $\Phi(m) = {\rm{d}}N/{\rm{d}}M$ is the black hole mass function. The results are presented in Table~\ref{tab:NM}.

\begin{table}
 \caption{The number density ($N_0$) and mass density ($M_0$) of supermassive black holes for mass functions shown in Figure \ref{fig:BHMF}.}
 \label{tab:NM}
 \begin{tabular}{lcc}
  \hline
  \multirow{2}{*}{Model} & $N_0$ & $M_0$ \\
    & $[10^{-3} {\rm{Mpc}}^{-3}]$ & $[10^{5} M_{\sun} {\rm{Mpc}}^{-3}]$ \\
  \hline
  \citet{Marconi2004} & $2.33$ & $1.86$ \\[2pt]
  \citet{Li2011} & $6.35$ & $4.70$ \\[2pt]
  \citet{Shankar2013} & $4.17$ & $6.01$ \\[2pt]
  \citet{Ueda2014} & $10.56$ & $9.48$ \\[2pt]
  \citet{MP2016} & $2.29$ & $3.18$ \\[2pt]
  \hline
 \end{tabular}
\end{table}

\bsp	
\label{lastpage}
\end{document}